\documentclass[conference]{IEEEtran}

\IEEEoverridecommandlockouts
\usepackage{cite}
\usepackage{amsmath,amssymb,amsfonts}
\usepackage{algorithmic}
\usepackage{graphicx}
\usepackage{textcomp}
\usepackage{xcolor}
\DeclareMathOperator*{\argmin}{arg\,min}
\def\BibTeX{{\rm B\kern-.05em{\sc i\kern-.025em b}\kern-.08em
    T\kern-.1667em\lower.7ex\hbox{E}\kern-.125emX}}

\begin{document}

\title{Deep-Learning based Motion Correction for Myocardial T\textsubscript{1} Mapping\textsuperscript{1}}

\author{\IEEEauthorblockN{Dar Arava, Mohammad Masarwy}
\IEEEauthorblockA{Andrew and Erna Viterbi Faculty of \\
Electrical and Computer Engineering \\
Technion – Israel Institute of Technology\\
Haifa, Israel \\
\{aravadar,mohammad.m\}@campus.technion.ac.il}
\and
\IEEEauthorblockN{Samah Khawaled \IEEEauthorrefmark{1}, Moti Freiman\IEEEauthorrefmark{2}}
\IEEEauthorblockA{\IEEEauthorrefmark{1} Department of Applied Mathematics \\
\IEEEauthorrefmark{2} Department of Biomedical Engineering \\
Technion – Israel Institute of Technology\\
 Haifa, Israel \\
ssamahkh@campus.technion.ac.il, moti.freiman@bm.technion.ac.il}
%\and 
%\IEEEauthorblockN{Moti Freiman}
%\IEEEauthorblockA{Department of Biomedical Engineering, \\
%Technion – Israel Institute of Technology\\
%Haifa, Israel \\
%moti.freiman@bm.technion.ac.il}

}

\maketitle
\addtocounter{footnote}{1}
\footnotetext{Accepted for presentation at the 2021  IEEE International Conference on Microwaves, Antennas, Communications and Electronic Systems (COMCAS)}
\begin{abstract}

Myocardial T\textsubscript{1} mapping is a cardiac MRI technique, used to assess myocardial fibrosis. In this technique, a series of T\textsubscript{1}-weighted MRI images are acquired with different saturation or inversion times. These images are fitted to the T\textsubscript{1} model to estimate the model parameters and construct the desired T\textsubscript{1} maps. In the presence of motion, the different T\textsubscript{1}-weighted images are not aligned. This, in turn, will cause errors and inaccuracies in the final estimation of the T\textsubscript{1} maps. Therefore, motion correction is a necessary process for myocardial T\textsubscript{1} mapping. 
We present a deep-learning (DL) based system for cardiac T\textsubscript{1}-weighted MRI images motion correction. When applying our DL-based motion correction system we achieve a statistically significant improved performance by means of R\textsuperscript{2} of the model fitting regression, in compared to the model fitting regression without motion correction ($0.52$ vs $0.29$, p$<$0.05).

\end{abstract}
\begin{IEEEkeywords}
Cardiac MRI, Deep Learning, Registration, Motion Correction,Myocardial T1 mapping, T1 Maps
\end{IEEEkeywords}

\section{Introduction}
Myocardial T\textsubscript{1} mapping is a cardiac magnetic resonance imaging (MRI) technique, which shows early clinical promise \cite{b10} in the assessment of cardiac fibrosis and inflammation by characterizing changes in the myocardial extracellular water (edema, focal, or diffuse fibrosis), fat, iron, and amyloid protein content \cite{b11}.
This imaging technique uses a series of T\textsubscript{1}-weighted images acquired with different saturation or inversion times. The intensity of the T\textsubscript{1}-weighted images behaves according to the following signal decay model \cite{b3}:
\begin{equation} \label{eu_eqn}
I = M_0(1 - e^{- \frac{t} {T_1}})
\end{equation}
where T\textsubscript{1} and $M_0$ are the model parameters. $M_0$ and the native T\textsubscript{1} maps are later estimated by a voxel-wise fitting of the aforementioned intensity signal decay model \eqref{eu_eqn}.

To prevent motion disruption while acquiring MRI images for T\textsubscript{1} mapping, patients are often requested to hold-breath during the scan\cite{b1,b2}. There have been attempts for reconstruction of T\textsubscript{1} maps from free-breathing sequences, for example by using a slice tracking respiratory navigator \cite{b3}, however, these methods still suffer from motion artifacts (caused by respiratory drifting or limitations of the prospective slice-tracking technique to reach full registration).

Respiratory and cardiac motion cause voxel misalignment between the different T\textsubscript{1}-weighted images. This, in turn, will cause errors, artifacts, and inaccuracies in the estimation of T\textsubscript{1} maps. Therefore, motion correction is essential in myocardial T\textsubscript{1} mapping. T\textsubscript{1}-weighted images should be aligned by a registration technique prior to performing the T\textsubscript{1} mapping. Post-processing registration methods find the deformation field that map between a pair of fixed and moving images into a combined coordinate system. Then, they register the moving image by warping it with the corresponding deformation field thereby reducing motion artifacts and misalignment between the two images.

 In \cite{b4}, a method of simulating synthetic free-motion T\textsubscript{1} images according to an initial noisy T\textsubscript{1} map is proposed. This synthetic series of images is used for the estimation of the deformation field by comparing it with the series of original T\textsubscript{1} images (with motion). This approach does not consider the T\textsubscript{1} model variations between patients. In \cite{b5}, a modified optical flow energy function is defined, including an additional term meant to prevent the formation of transient structures from through-plane motions. In this approach, estimated motion parameters were effected by signal-to-noise and contrast-to-noise ratios of the acquired images. Both of the aforementioned methods \cite{b4,b5} were performed on breath-holding acquisitions(which minimize the effect of motion) and require high computation resources. In \cite{b6}, a strategy of non-rigid registration of free-breathing T\textsubscript{1} images based on active shape models is proposed. This method, however, requires a new optimization calculation of the motion deformation for each new previously unseen data input at inference phase.

To tackle this obstacle, we propose a deep-learning (DL) approach for the motion correction of T\textsubscript{1} images. Utilization of a DL-based system in such post-processing registration tasks enables a faster registration, since once the network is trained, the deformation field is estimated by feed-forwarding the pair of fixed and moving images into the network \cite{b7,b9}. Unlike classical methods, solving an optimization problem for an unseen input of images is not needed at inference time. 
Additionally, we employ an unsupervised DL system to obtain a free-form deformation field, which is not limited to a certain model and can characterize the deformations caused by the cardiac motion. In contrast to supervised schemes, unsupervised registration models ``learn'' the transformation that maps from moving to fixed images without providing a ground truth about the transformation. 
Our motion correction system is based on the unsupervised DL model of \texttt{voxelmorph} \cite{b7}. Our model was trained and evaluated on myocardial T\textsubscript{1}-weighted images, which were acquired with a free-breathing scan. 

\begin{figure}[t]
\centerline{\includegraphics[width=9cm]{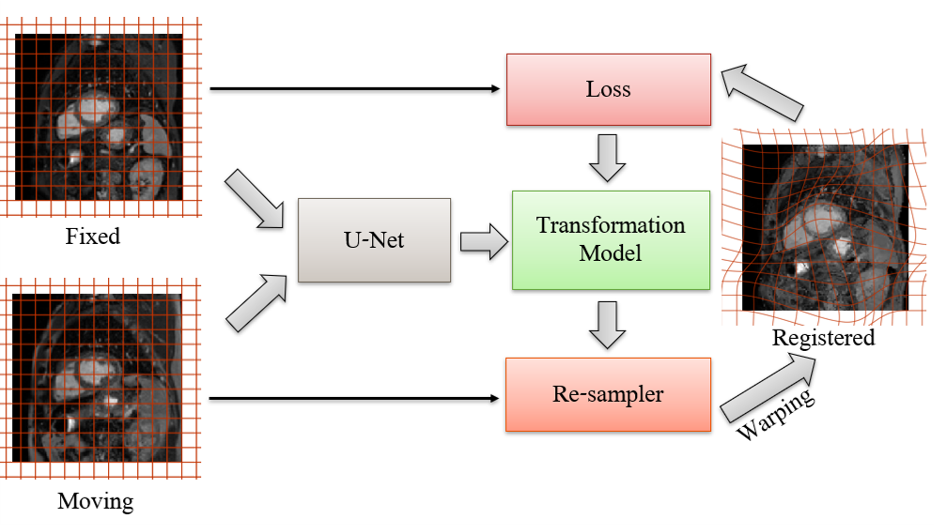}}
\caption{\textbf{Unsupervised DL-based registration system.} The inputs of the DL model are 3D cardiac MRI images acquired at consecutive times. Fixed and moving images are fed as two-channel input to U-net. The U-net predicts, as an output, the 3D deformation field that map between this pair of images. In the training phase, the loss term is computed between the fixed image and the resulting registered image.}
\label{fig:system}
\end{figure}

\section{Motion correction: proposed method}
Deformable registration can be formulated as an optimization problem. Let us denote the pair of fixed and moving images by $I_{f}$ and $I_M$, respectively. $\Phi$ is the deformable transformation, which accounts for mapping the grid of $I_M$ to the grid of $I_f$. Then, the energy functional that we aim to optimize is:
\begin{equation}
\argmin_\Phi {S(I_f,I_M\circ\Phi)+\lambda R(\Phi)} \label{eq:funcreg1}
\end{equation}
where $I_M\circ\Phi$ denotes the result of warping the moving image with $\Phi$. $S$ is a dissimilarity metric which quantifies the resemblance between the resulted image and the fixed input, and $R$ is a regularization term that penalizes the deformation smoothness. The scalar $\lambda$ is a tuning parameter that accounts for balancing between the two terms and it controls the smoothness of the resulting deformation. Popular dissimilarity metrics are the mean squared difference/error (MSE), the cross-correlation, and the mutual information (MI). We use the negative MI, as a dissimilarity loss between the resulted registered warped image and the fixed image. Local\footnote{local is advantageous over global since it preserves spatial relations and is more efficient computationally} MI is empirically calculated on blocks of size $16\times16\times4$
\begin{equation}
S(I_f,I_M\circ\Phi)=-\sum_{x\in I_f} \sum_{y\in I_M\circ\Phi} p(x,y)log\left(\frac{p(x,y)}{p(x)p(y)}\right)\label{eq1}
\end{equation}
As opposed to other common loss functions, such as the MSE, the MI preserves the natural image contrast change over time, which better suits the  T\textsubscript{1} decay model mentioned in \eqref{eu_eqn}. 
Additionally, we set the regularization term to be equal to the l\textsubscript{2} norm of the deformation gradients:
\begin{equation}
R(\Phi) = \sqrt{\sum_{i,j}|\nabla \Phi(i,j)|^2}\label{eq2}
\end{equation}
In DL-based registration, the task of the model is to predict the deformation: $\hat{\Phi} = f_\theta(I_{M},I_{F})$, where $\theta$ are the parameters of the network and $I_{M},I_{F}$ are the input images. Then, the network predicts the deformation, $\Phi$, by optimizing the following: 
\begin{equation}
\hat{\Phi}= \argmin_\theta {S(I_F,I_M\circ f_\theta(I_{M},I_{F}))+\lambda R(f_\theta(I_{M},I_{F}))} \label{eq:funcreg2}
\end{equation}   
The scheme of our registration system is given in Fig.~\ref{fig:system}. Given a pair of moving ($I_M$) and target ($I_F$) images as a 2-channel input, the model predicts the deformation field, $\Phi=f_\theta(I_{M},I_{F})$. The deformation field is obtained by a pretrained U-Net model. Lastly, it maps each voxel, $p$, in the moving image to $\Phi(p)$ by applying the linear spatial interpolation according the corresponding deformation field.
\begin{figure}[t]
\centerline{\includegraphics[width=7.1cm]{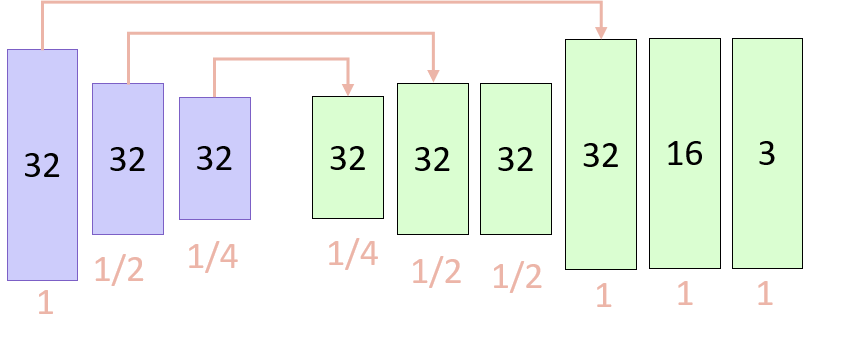}}
\caption{\textbf{The U-Net architecture.} The U-net model consists of three encoder layers (blue) and six decoder layers (green) with skip connection from the encoder layers to the decoder layers (highlighted by the corresponding arrows). The last layer in the decoder is 3-channel convolution layer that outputs the 3-directional deformation field.}
\label{fig:arch}
\end{figure}
\paragraph*{Network architecture}
Our main building block is a U-Net based model, consisting of encoder and decoder layers with skip connections. The architecture of the U-Net is based on the \texttt{VoxelMorph} model \cite{b7}. Both the encoder and decoder layers consist of convolutional neural network (CNN) layers with kernel size $3\times3\times 3$ followed by Leaky ReLU activation functions. The U-Net includes propagation of features learned in the encoder stage to the decoder layers, with the number of hidden layers as highlighted in Fig.~\ref{fig:arch}. The input of the U-Net is a pair of 3D fixed and moving images taken at consecutive inverse times.  The output of the U-Net is a 3 directional deformation field used in warping the moving image. The implementation of our registration network is based on an open source model of \texttt{VoxelMorph} \cite{b7}.

\begin{figure}[t]
\centerline{\includegraphics[scale=0.35]{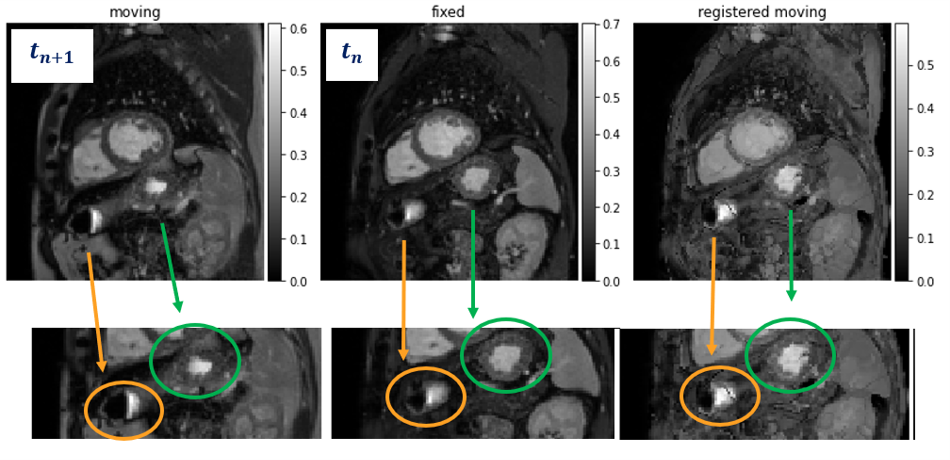}}
\caption{\textbf{Example of the registration result.} From left to right: a selected slice from the 3D moving, fixed, and the resulting registered (warped) image. The bottom line shows highlighted and enlarged areas of the three selected images. The warped resulting image exhibits noticeable non-rigid movements (see the shapes marked by green and orange circles). }
\label{fig:regexample}
\end{figure}
\section{Experiments and Results}
\subsection{Database and Preprocessing}
The data set used in this work is adopted from \cite{b6}. This data set consists of 3D free-breathing T\textsubscript{1}-weighted images from 210 subjects. Cardiac MRI images are of size $320\times 320 \times 5$, where the third dimension consists of five short axial slices located along the left ventricular. For each slice location, 11 T\textsubscript{1}-weighted images were captured at the time points:  (TI\textsubscript{1}, TI\textsubscript{1}+RR, TI\textsubscript{1}+2RR, …TI\textsubscript{1}+4RR, TI\textsubscript{2}, TI\textsubscript{2}+RR, ..., TI\textsubscript{2}+4RR, $\infty$). The initial inversion times TI\textsubscript{1}, TI\textsubscript{2} are set to $135[ms]$, $350 [ms]$ and RR is the duration of one heart-beat \cite{b3}. Further details about the T\textsubscript{1} images can be found in \cite{b6}.

Images were initially cropped to the central region of interests (ROIs). Then, they were down-sampled and zero-padded, in z dimension only, to be of size $128\times 128 \times 8$. Cropping and down-sampling in x and y dimensions were applied to decrease the calculation complexity and for adjusting the size of the U-Net inputs to be of powers of $2$. In addition, images were normalized to gray levels within the range $[0,1]$.
Lastly, the dataset was randomly divided into training and test sets with ratios of $80\%$ and $20\%$, respectively. 

\subsection{Results}
We performed motion correction for image pairs with consecutive inverse times. Fig.~\ref{fig:regexample} depicts the registration results. The resulting warped image has a resemblance in both shape and coordinate system to that of the fixed image. Further, the motion correction preserved the contrast and intensities of the moving image. 

\begin{figure}[t]
\centerline{\includegraphics[scale=0.35]{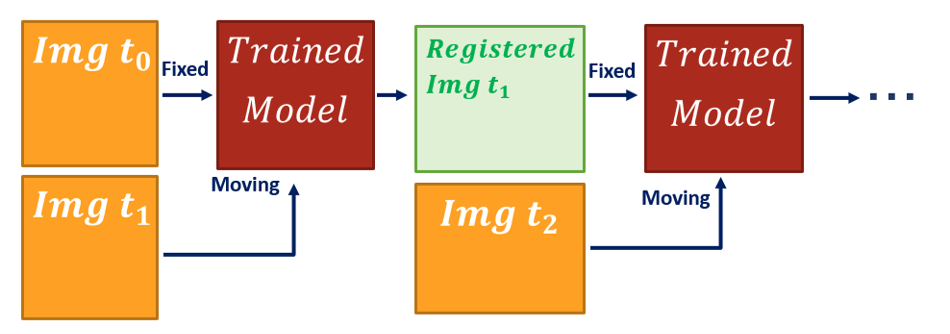}}
\caption{\textbf{Cascading scheme for applying the pair-wise registration model on the whole series of images.}  We first align between the first two images at time $t=0$ and $t=1$. The output warped image is then used as a fixed image for the next registration block, where the moving input is the image at time $t=2$. This process is repeated for all time points to achieve alignment for the whole set of images in the series.}
\label{fig4}
\end{figure}

To align between the series of  T\textsubscript{1}-weighted images of all 11 time points, we cascaded the registration systems recursively. Fig.~\ref{fig4} shows the cascading scheme for our model. Firstly, we register between T\textsubscript{1}-weighted images of times $t=0$ and $t=1$. Then, registration is performed in a recursive manner by forwarding the registered image from the block $n$ as a fixed image input to the next registration unit, where the image of time $t=n+1$ is used as the moving input.

\begin{figure}[t]
\centerline{\includegraphics[scale=0.35]{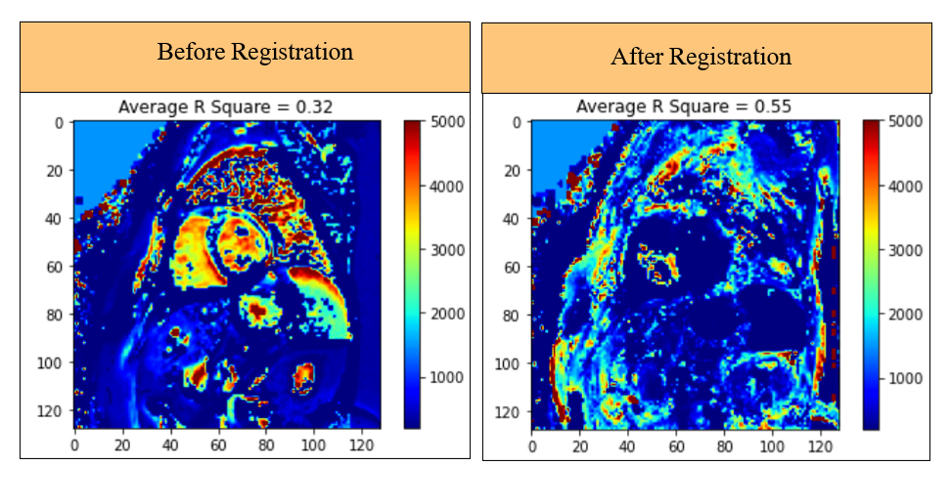}}
\caption{\textbf{T\textsubscript{1} mapping before and after motion correction.} A slice from the 3D voxel-wise T\textsubscript{1} map estimated from the T\textsubscript{1}-weighted images before and after registration. For each voxel, the intensity  levels are fitted to the T\textsubscript{1} decay model. The averaged R\textsuperscript{2} of the regression, calculated over the whole voxels, is annotated. We notice an improvement in terms of R\textsuperscript{2} after performing motion correction.}
\label{fig5}
\end{figure}

\begin{figure}[t]
\centerline{\includegraphics[scale=0.35]{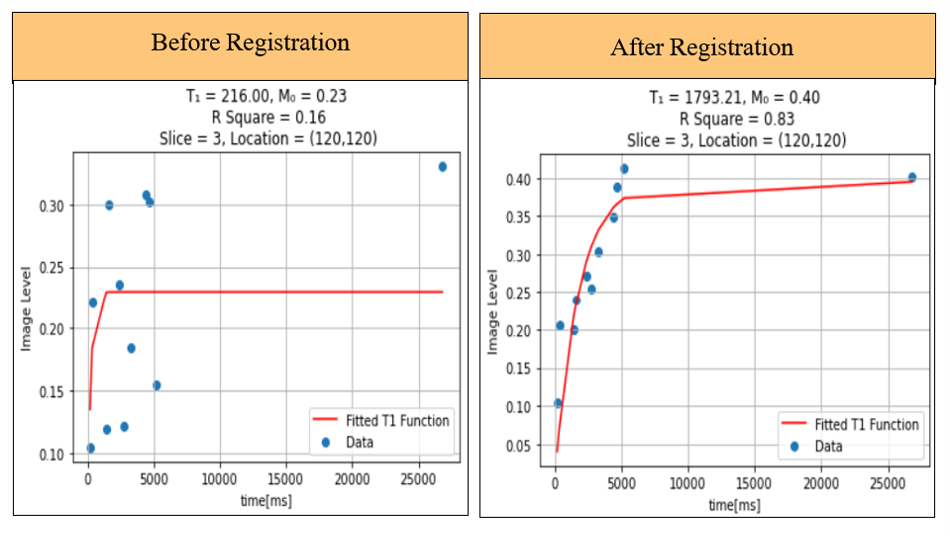}}
\caption{\textbf{Scatter plots of the voxel-wise regression to T\textsubscript{1} decay model, before and after motion correction.} Motion correction improves the fitting to the model and better regression results are achieved ( higher R\textsuperscript{2}).}
\label{fig6}
\end{figure}

Lastly, after having the series of T\textsubscript{1}-weighted images aligned, we perform quantitative T\textsubscript{1} mapping. T\textsubscript{1} maps are constructed by a voxel-wise fitting of a signal-decay model \eqref{eu_eqn}. We investigate the effect of motion correction on the resulting  T\textsubscript{1} maps by comparing the estimated regression results before and after applying the alignment of the T\textsubscript{1}-weighted images. We noticed improvement in fitting to the T\textsubscript{1} decay model, as we achieved an increased R\textsuperscript{2} of the regression after registration. The average R\textsuperscript{2} and the standard deviation (std) is calculated over the whole test set. We obtained mean R\textsuperscript{2} of $0.52$ with std of $0.09$ after applying motion correction. However, estimation T\textsubscript{1} without motion correction yields mean R\textsuperscript{2} of $0.29$ with standard deviation of $0.06$. Fig.~\ref{fig5} illustrates an example of T\textsubscript{1} mapping results before and after applying motion correction. The fitting of the intensity levels of the images to the signal-decay model are performed per voxel for each of the images in the test set. Voxel-wise scatter plots of the regression and the T\textsubscript{1} decay model before and after motion correction are presented in Fig.~\ref{fig6}. 

\begin{figure}[t]
\centerline{\includegraphics[scale=0.4]{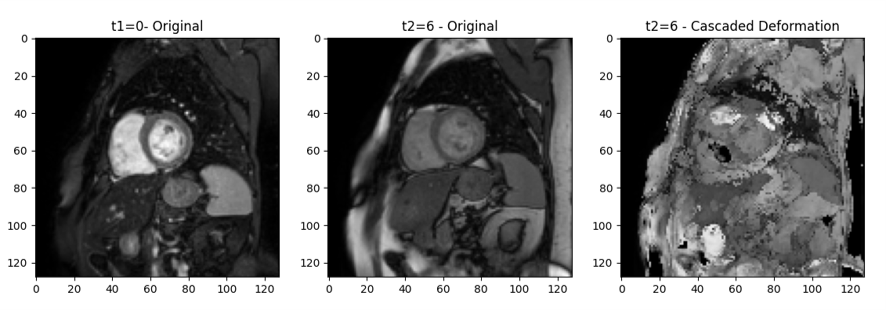}}
\caption{\textbf{Example of distorted image after cascading the model six times.} From left to right:  T\textsubscript{1}-weighted  image at time $t=0$, which is used as the input moving image, the fixed image (at time $t=6$) and the resulted warped after cascading the registration blocks six times. }
\label{fig7}
\end{figure}

\begin{figure}[t]
\centerline{\includegraphics[scale=0.4]{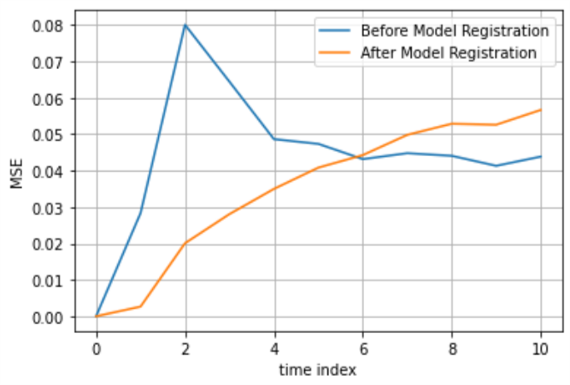}}
\caption{\textbf{MSE of registration over time, before and after motion correction.} The MSE is calculated between a reference image taken from the T\textsubscript{1}-weighted images at time $t=0$ and each one of the other images at times $t=1,..,10$. At initial time points, we notice an improvement in MSE after performing motion correction (orange curve), compared with that obtained without motion correction (blue curve). However, as a result of the accumulated error in the cascaded scheme the MSE increases starting from $t=6$. }
\label{fig8}
\end{figure}

Although we achieved an increased R\textsuperscript{2} in the fitting to the T\textsubscript{1} decay model after registration, there is a place for improvement in the performance of fitting. This slight improvement (R\textsuperscript{2} is smaller than $0.95$ , Fig.~\ref{fig5}) is caused by artificial distortion in the resulting registered image that were incorporated during cascading of the registration systems for $11$ times. In fact, cascading registration models causes accumulating of registration error that lead to a considerable visible artifacts to the final warped image (See Fig.~\ref{fig7}). To inspect this effect, the MSE between the fixed image, image at time $t=0$, to each one of the other images in times $t=1,...,10$ is calculated. At initial time point, motion correction leads to lower values of  MSE, compared to MSE before registration. However, at a certain time point ($t=6$) accumulating registration error and distortions lead to higher MSE values (See Fig.~\ref{fig8}).

Another attempt to extend the motion correction model to deal with larger set of images is to train the network on image pairs consisting of images at t\textsubscript{0} (as the fixed image) and another randomly selected image from times  $t=1,...,10$. This training fashion enables estimation of deformation fields that characterize larger movements.  However, due to the intensity decaying behavior and gray-level variation between images with large time ranges, this led to visible artifacts in the registered image. 

\section{Conclusions}
In this paper, we performed a motion-correction for cardiac T\textsubscript{1}-weighted images for quantitative T\textsubscript{1} mapping. Our motion correction system based on the unsupervised DL registration system, \texttt{VoxelMorph} \cite{b7}. T\textsubscript{1} maps are reconstructed by voxel-wise signal-decay model fitting. Our DL-based motion correction system shows an improved performance in the estimation of T\textsubscript{1} maps, as it provides better model fitting in terms of R\textsuperscript{2} of the regression.

A simple cascading approach of our trained model showed a slight improvement in the fit to the  T\textsubscript{1} model, however, it led to artifacts in the final registered image due to the registration error that accumulated during cascading of the units. To overcome this challenge, one may incorporate the T\textsubscript{1} decaying model into the loss of the DL network, and the whole series of T\textsubscript{1}-weighted images should be utilized as an input rater that only a pair of fixed and moving images.

\section*{acknowledgments}
Khawaled, S. is a fellow of the Ariane de Rothschild
Women Doctoral Program. Freiman, M. is a Taub fellow
(supported by the Taub Family Foundation, Technion’s program for
leaders in Science and Technology). We wish to thank Mr. Johanan Erez, Ms. Ina Talmon and the Vision and Image Sciences Laboratory (VISL) 
staff for their ongoing assistance, support, and providing technical tools for this project.    

.
\vspace{12pt}

\begin{thebibliography}{00}
\bibitem{b10} Christine L. Jellis, Deborah H. Kwon. Myocardial T1 mapping: modalities and clinical applications. Cardiovasc Diagn Ther. 2014 Apr; 4(2): 126–137. doi: 10.3978/j.issn.2223-3652.2013.09.03.
\bibitem{b11} Dina Radenkovic, Sebastian Weingärtner, Lewis Ricketts, James C. Moon, Gabriella Captur. T1 mapping in cardiac MRI. Heart Fail Rev. 2017; 22(4): 415–430. Published online 2017 Jun 16. doi: 10.1007/s10741-017-9627-2.
\bibitem{b3} Weing€artner S, Roujol S, Akc¸akaya M, Basha TA, Nezafat R. Freebreathing multislice native myocardial T1 mapping using the sliceinterleaved
T1 (STONE) sequence. Magn Reson Med 2015;74:115–124.
\bibitem{b1} Piechnik SK, Ferreira VM, Dall’Armellina E, Cochlin LE, Greiser A, Neubauer S, Robson MD. Shortened Modified Look-Locker Inversion
recovery (ShMOLLI) for clinical myocardial T1-mapping at 1.5 and 3 T within a 9 heartbeat breathhold. J Cardiovasc Magn Reson 2010; 12:69.
\bibitem{b2} Marks B, Mitchell DG, Simelaro JP. Breath-holding in healthy and pulmonary-compromised populations: effects of hyperventilation and oxygen inspiration. J Magn Reson Imaging 7:595–597.
\bibitem{b4} Xue H, Shah S, Greiser A, Guetter C, Littmann A, Jolly M-P, Arai AE, Zuehlsdorff S, Guehring J, Kellman P. Motion correction for myocardial T1 mapping using image registration with synthetic image estimation. Magn Reson Med 2012;67:1644–1655.
\bibitem{b5} Roujol S, Foppa M, Weing€artner S, Manning WJ, Nezafat R. Adaptive registration of varying contrast-weighted images for improved tissue characterization (ARCTIC): application to T1 mapping. Magn Reson Med 2015;73:1469–1482.
\bibitem{b6} Hossam El-Rewaidy, Maryam Nezafat, Jihye Jang, Shiro Nakamori, Ahmed S. Fahmy, and Reza Nezafat. Nonrigid Active Shape Model–Based Registration Framework for Motion Correction of Cardiac T1 Mapping. Magnetic Resonance in Medicine 2018;80:780–791
\bibitem{b7} Guha Balakrishnan, Amy Zhao, Mert R. Sabuncu, John Guttag, Adrian V. Dalca. An Unsupervised Learning Model for Deformable Medical Image Registration. Proceedings of the IEEE Conference on Computer Vision and Pattern Recognition (CVPR), 2018;9252-9260
\bibitem{b8} Guo, Courtney K. Multi-modal image registration with unsupervised deep learning. MEng. Thesis 
\bibitem{b9} Unsupervised Learning of Probabilistic Diffeomorphic Registration for Images and Surfaces
Adrian V. Dalca, Guha Balakrishnan, John Guttag, Mert R. Sabuncu MedIA: Medial Image Analysis. 2019. eprint arXiv:1903.03545


\end{thebibliography}
\end{document}